\documentclass[twocolumn,showpacs,prb,amsfonts,amsmath,amssymb,floatfix]{revtex4}
\usepackage{color}
\usepackage{hhline}
\usepackage{mathrsfs}
\usepackage{graphicx}
\usepackage{dcolumn}
\usepackage{bm}
\usepackage{multirow}
\usepackage{booktabs}
\usepackage{afterpage}
\usepackage{amsmath}
\usepackage{multirow}

\begin{document}

\title{{\it Ab initio} $GW$ plus cumulant calculation for isolated band system: Application to organic conductor (TMTSF)$_2$PF$_6$ and transition-metal oxide SrVO$_3$} 

\author{Kazuma Nakamura$^{1}$}
\author{Yoshiro Nohara$^{2}$} 
\author{Yoshihide Yoshimoto$^{3}$}
\author{Yusuke Nomura$^{4}$} 
\affiliation{$^1$Quantum Physics Section, Kyushu Institute of Technology, 1-1 Sensui-cho, Tobata, Kitakyushu, Fukuoka, 804-8550, Japan}
\affiliation{$^2$Max-Planck-Institute for Solid State Research, Heisenbergstr. 1, D-70569 Stuttgart, Germany} 
\affiliation{$^3$Department of Computer Science, The University of Tokyo, 7-3-1 Hongo, Bunkyo-ku, Tokyo 113-0033, Japan}
\affiliation{$^4$Centre de Physique Th\'eorique, \'Ecole polytechnique, CNRS, Universit\'e Paris-Saclay, F-91128 Palaiseau, France} 

\date{\today}
\begin{abstract}
We present {\em ab initio} $GW$ plus cumulant-expansion calculations for an organic compound (TMTSF)$_2$PF$_6$ and a transition-metal oxide SrVO$_3$. These materials exhibit characteristic low-energy band structures around the Fermi level, which bring about interesting low-energy properties; the low-energy bands near the Fermi level are isolated from the other bands and, in the isolated bands, unusually low-energy plasmon excitations occur. To study the effect of this low-energy-plasmon fluctuation on the electronic structure, we calculate spectral functions and photoemission spectra using the {\em ab initio} cumulant expansion of the Green's function based on the $GW$ self-energy. We found that the low-energy plasmon fluctuation leads to an appreciable renormalization of the low-energy bands and a transfer of the spectral weight into the incoherent part, thus resulting in an agreement with experimental photoemission data.

\end{abstract}


\maketitle

\section{Introduction}\label{sec:intro} 
The understanding of low-energy electronic structures and excitations in real materials is an important subject of condensed-matter physics and material science. Interesting phenomena such as a non-Fermi-liquid behavior and unconventional superconductivity are caused by the instability of electronic structures near the Fermi level. A common feature is often found in the band structures showing such phenomena; isolated bands appear near the Fermi level. The width of these isolated bands is typically the order of 1 eV, which is comparable to local electronic interactions. Thus, in these isolated bands, the kinetic and potential energies compete with each other, and the competition is often discussed within a local-interaction approximation, as in the Hubbard model. In real materials, however, there exist various elementally excitations not described by the local electronic interaction. The plasmon in metallic systems or the exciton in insulating systems are well known examples of such nonlocal excitations, which result from the long-range Coulomb interaction. In the above-mentioned isolated-band systems, the plasmon excitation can occur in this band, and its energy scale can be very small (of the order of 1 eV), which is comparable to the bandwidth and the size of the local Coulomb interaction. 

In this study, we investigate the effect of the low-energy-plasmon fluctuation on the electronic structure of real isolated-band systems from first principles. For this purpose, we choose two materials, a quasi-one dimensional organic conductor (TMTSF)$_2$PF$_6$ (Ref.~\onlinecite{Bechgaard,Ishiguro-Yamaji-Saito,Kuroki,Ishibashi}), where TMTSF stands for tetramethyltetraselenafulvalene, and a three-dimensional perovskite transition-metal oxide SrVO$_3$ (Ref.~\onlinecite{LDA+DMFT-1, LDA+DMFT-2, LDA+DMFT-3}). These materials are typical isolated-band systems and are studied as benchmark materials of the correlated metal, where the local Coulomb-interaction effect on the electronic properties is investigated with much interest.~\cite{Kuroki,LDA+DMFT-1, LDA+DMFT-2, LDA+DMFT-3} In the present work, we focus on the low-energy plasmon effect on the electronic structure.~\cite{TMTSF-plasmon-Nakamura,Gatti} Through the comparison between theoretical and experimental results on the plasmon-related properties and spectral functions, we verify low-energy plasmon effects on the electronic structure of the real system. 

The organic conductor (TMTSF)$_2$PF$_6$ is a representative quasi-one-dimensional material,~\cite{Bechgaard,Ishiguro-Yamaji-Saito,structure-TMTSF} and basically behaves as a good metallic conductor.~\cite{TMTSF-R-Dressel,TMTSF-R-Jacobsen} At low temperature (around 12 K), it undergoes a transition to a spin-density-wave phase.~\cite{TMTSF-R-Dressel} In the high-temperature metallic region, photoemission spectroscopy has observed small spectral weight near the Fermi level.~\cite{TMTSF-PES-1,TMTSF-PES-2,TMTSF-PES-3,TMTSF-PES-4,TMTSF-PES-5} From this observation and the quasi-one-dimensional nature, the origin of the renormalization has been discussed in view of the Tomonaga-Luttinger liquid.~\cite{TMTSF-PES-2,TMTSF-PES-5}  On the other hand, this material exhibits clear low-energy plasma edges around 0.1-1 eV in the reflectance spectra.~\cite{TMTSF-R-Dressel,TMTSF-R-Jacobsen} Therefore, this plasmon excitation would also be a prominent renormalization factor of the electronic structure. 

The transition-metal oxide SrVO$_3$ is another well known correlated metal.~\cite{SVO-struct,R-SVO} Many high-resolution photoemission measurements~\cite{PES-SVO-60} including the bulk-sensitive version~\cite{PES-SVO-900,SVO-PES} were performed and clarified a strong renormalization of the low-energy isolated band and the satellite peak just below this band. The origin has actively been discussed in terms of the local electronic correlation.~\cite{LDA+DMFT-1, LDA+DMFT-2, LDA+DMFT-3, LDA+DMFT-4, LDA+DMFT-5, LDA+DMFT-6, LDA+DMFT-7, LDA+DMFT-8, GW+DMFT-1, GW+DMFT-2, GW+DMFT-3, GW+DMFT-4, GW+DMFT-5, GW+DMFT-6} On the other hand, the reflectance and electron-energy-loss spectra have clarified low-energy plasmon excitations around 1.4 eV.~\cite{R-SVO} In density-functional band structure, SrVO$_3$ has isolated bands of the $t_{2g}$ orbitals around the Fermi level, whose bandwidth is about 2.7 eV.~\cite{LDA+DMFT-1,LDA+DMFT-2,LDA+DMFT-3} Also, the constrained random phase approximation gives an estimate of the local electronic interaction of $\sim$2-3 eV.~\cite{cRPA-1,cRPA-2} Thus, the energy scale of the experimentally observed plasmon excitation is comparable to the bandwidth and the local Coulomb interaction. Hence, the low-energy plasmon fluctuation would certainly be relevant to the low-energy properties. 

To study how the plasmon excitation affects the electronic structure, we perform {\em ab initio} calculations based on the $GW$ approximation.~\cite{Hedin,Hybertsen,Onida,Aryasetiawan,GW-1,GW-2,GW-3,GW-4,GW-5,GW-6,GW-7,GW-8,GW-9,GW-10,GW-11,GW-12,GW-13,GW-14,GW-15,GW-16,GW-17,GW-18,GW-19,GW-20,Nohara,TMTSF-plasmon-Nakamura} The $GW$ calculation considers the self-energy effect due to the plasmon fluctuation and describes properly quasiparticle energies in the valence region. On the other hand, the description for the plasmon satellite in the spectral function is known to be less accurate.~\cite{Aryasetiawan} In order to improve this deficiency, {\em ab initio} $GW$ plus cumulant ($GW$+$C$) calculations have recently been performed.~\cite{Gatti,GWC-1,GWC-2,GWC-3,GWC-4,GWC-5,GWC-6,GWC-7} The accuracy of the $GW$+$C$ method has been verified in bulk silicon~\cite{GWC-3,GWC-4} and simple metals,~\cite{GWC-2} where the satellite property is satisfactorily improved. In the present study, we apply the $GW$+$C$ method to the study of the above mentioned isolated-band systems, and show that the low-energy plasmon fluctuations modify substantially the low-energy electronic structure. 

The present paper is organized as follows. In Sec. II, we describe the $GW$ and $GW$+$C$ methods to calculate dielectric and spectral properties. Computational details and results for (TMTSF)$_2$PF$_6$ and SrVO$_3$ are given in Sec. III. We also discuss the comparison between theory and experiment, focusing on the renormalization of the electronic structure due to the plasmon excitation. The summary is given in Sec. IV. 

\section{Method}\label{sec:method}

In this section, we describe {\em ab initio} $GW$ and $GW+C$ methods. The latter is a post-$GW$ treatment and uses the self-energies calculated with the $GW$ approximation. Below, we first describe details of the $GW$ calculation. 
\subsection{GW approximation} 
The non-interacting Green's function is written as  
\begin{eqnarray}
G_0({\bf r,r'},\omega) = \sum_{\alpha{\bf k}} \frac{\phi_{\alpha{\bf k}}({\bf r}) \phi^{*}_{\alpha{\bf k}}({\bf r'})} {\omega-\epsilon_{\alpha{\bf k}}+i \delta {\rm sgn}(\epsilon_{\alpha{\bf k}}-\mu)},   
\end{eqnarray}
where ${\bf k}$ and $\omega$ are wavevector and frequency, respectively. $\phi_{\alpha{\bf k}}({\bf r})$ and $\epsilon_{\alpha{\bf k}}$ are the Kohn-Sham (KS) wavefunction and its energy, respectively, and $\mu$ is the Fermi level of the KS system. The $\delta$ parameter is chosen to be a small positive value to stabilize numerical calculations.

A polarization function of a type $-i G_0 G_0$ is written in a matrix form in the plane wave basis as
\begin{eqnarray}
\chi_{{\bf GG'}}({\bf q},\omega)&=&2\sum_{{\bf k}}\sum^{unocc}_{\alpha}\sum^{occ}_{\beta} M_{\alpha\beta}^{{\bf G}}({\bf k+q,k}) \nonumber \\
&\times&M_{\alpha\beta}^{{\bf G'}}({\bf k+q,k})^{*} X_{\alpha{\bf k+q},\beta{\bf k}}(\omega)  \label{eq:chi} 
\end{eqnarray} 
with $M_{\alpha\beta}^{{\bf G}}({\bf k+q,k})=\langle\phi_{\alpha{\bf k+q}}|e^{i({\bf q+G})\cdot{\bf r}}|\phi_{\beta{\bf k}}\rangle$ and 
\begin{eqnarray}
X_{\alpha{\bf k+q},\beta{\bf k}}(\omega)\!=\! \frac{1}{\omega\!-\!\epsilon_{\alpha{\bf k\!+\!q}}\!+\!\epsilon_{\beta{\bf k}}\!+\!i\delta}\!-\!
\frac{1}{\omega\!+\!\epsilon_{\alpha{\bf k\!+\!q}}\!-\!\epsilon_{\beta{\bf k}}\!-\!i\delta}, \nonumber \\  
\end{eqnarray}
where ${\bf G}$ is a reciprocal lattice vector. With this polarization function, the symmetrized dielectric matrix in reciprocal space is defined by 
\begin{eqnarray}\
\epsilon_{{\bf G\!G'}}\!({\bf q},\omega)\!=\!\delta_{{\bf G\!G'}}\!-\!\frac{4\pi}{V}\frac{1}{|{\bf q\!+\!G}|}\chi_{{\bf G\!G'}}({\bf q},\!\omega)\frac{1}{|{\bf q\!+\!G'}|}  \label{symeps}
\end{eqnarray}
with $V$ being the crystal volume. In the ${\bf q+G}\to0$ limit, the dielectric matrix in Eq.~(\ref{symeps}) is expressed by~\cite{Hybertsen-RPA} 
\begin{widetext} 
\begin{eqnarray}
  \epsilon_{{\bf GG'}}(0,\omega) 
  = \left\{
    \begin{array}{@{\,\,\,\,\,\,\,}l@{\,\,\,\,\,\,\,\,\,\,\,\,\,\,\,}l}
     \displaystyle
     1-\frac{4\pi}{V} 2\sum_{{\bf k}}\sum^{unocc}_{\alpha}\sum^{occ}_{\beta}  
     \Biggl|\frac{p^{\mu}_{\alpha\beta}({\bf k})}
     {\epsilon_{\alpha{\bf k}}-\epsilon_{\beta{\bf k}}}\Biggr|^2 
     X_{\alpha{\bf k},\beta{\bf k}}(\omega)     
     -\frac{(\omega_{pl,\mu\mu})^2}{\omega(\omega+i\delta)}
     & \mbox{(${\bf G}={\bf G'}={\bf 0}$),}
     \\[+20pt]
     \displaystyle 
     -\frac{4\pi}{V} 
     2\sum_{{\bf k}}\sum^{unocc}_{\alpha}\sum^{occ}_{\beta} 
     \Biggl(\frac{p^{\mu}_{\alpha\beta}({\bf k})}
     {\epsilon_{\alpha{\bf k}}-\epsilon_{\beta{\bf k}}}\Biggr)
     \frac{M_{\alpha\beta}^{{\bf G'}}({\bf k,k})^*}{|{\bf G'}|}  
     X_{\alpha{\bf k},\beta{\bf k}}(\omega) 
     & \mbox{(${\bf G=0},{\bf G'\ne 0}$),} 
     \\[+20pt]
     \displaystyle 
     -\frac{4\pi}{V}
     2\sum_{{\bf k}}\sum^{unocc}_{\alpha}\sum^{occ}_{\beta} 
     \frac{M_{\alpha\beta}^{{\bf G}}({\bf k,k})}{|{\bf G}|}  
     \Biggl(\frac{p^{\mu}_{\alpha\beta}({\bf k})}
     {\epsilon_{\alpha{\bf k}}-\epsilon_{\beta{\bf k}}}\Biggr)^* 
     X_{\alpha{\bf k},\beta{\bf k}}(\omega) 
     & \mbox{($ {\bf G \ne 0}, {\bf G'=0}$),} 
    \\[+20pt]
     \displaystyle
     \delta_{{\bf GG'}}-\frac{4\pi}{V} \frac{1}{|{\bf G}|}
     \chi_{{\bf GG'}}({\bf 0},\omega)\frac{1}{|{\bf G'}|}
     & \mbox{($otherwise$)}
    \end{array}
  \right.
  \label{eq0}  
\end{eqnarray}
\end{widetext} 
where ${\bf q}$ approaches zero along the Cartesian $\mu$ direction and $p_{\alpha\beta{\bf k}}^{\mu}$ is a matrix element of a momentum as 
\begin{eqnarray}
p_{\alpha\beta{\bf k}}^{\mu}=-i\langle\phi_{\alpha{\bf k}}|\frac{\partial}{\partial x_{\mu}}+[V_{NL},x_{\mu}]|\phi_{\beta{\bf k}}\rangle,   
\label{p_ij} 
\end{eqnarray}
with $V_{NL}$ being the nonlocal part of the pseudopotential. In the first line of Eq.~(\ref{eq0}), the last term on the right hand side is the Drude term of the intraband transitions around the Fermi level,~\cite{Draxl-1,Draxl-2,GW-18} where 
\begin{eqnarray}
\omega_{pl,\mu\nu}=\sqrt{\frac{8\pi}{\Omega N} \sum_{\alpha{\bf k}} p_{\alpha\alpha{\bf k}}^{\mu} p_{\alpha\alpha{\bf k}}^{\nu} \delta(\epsilon_{\alpha {\bf k}}-\mu}) \label{wpl} 
\end{eqnarray} 
is the bare plasma frequency. The other terms in Eq.~(\ref{eq0}) result from the interband transitions.

We next describe the calculation of the self-energy. The operator of the exchange self-energy is defined by  
\begin{eqnarray}
\Sigma^X({\bf r,r'})\!=\!i\!\int \frac{d\omega}{2\pi} G_0({\bf r,r'},\omega) v({\bf r,r'}),    
\label{GW_SIGMAX}
\end{eqnarray}
 where $v({\bf r,r'})=1/|{\bf r-r'}|$ is the bare Coulomb interaction. In practice, we use an attenuation Coulomb interaction $\tilde{v}({\bf r,r'})=\Theta(R_c-|{\bf r-r'}|)/|{\bf r-r'}|$ with a cutoff $R_c$ instead of $v$ to treat the integrable singularities in the bare Coulomb interaction.~\cite{Spencer} The matrix element of the exchange self-energy is thus 
\begin{eqnarray} 
\Sigma^{X}_{\alpha\beta{\bf k}}
&=& \int\!d{\bf r} \int\!d{\bf r'}\!\phi_{\alpha{\bf k}}^*({\bf r})\Sigma^X({\bf r,r'})\phi_{\beta{\bf k}}({\bf r'}) \nonumber \\ 
&=& \Sigma^{X,body}_{\alpha\beta{\bf k}}+\Sigma^{X,head}_{\alpha\beta{\bf k}},  \label{Sxb+Sxh} 
\end{eqnarray}
where $\Sigma^{X,body}$ and $\Sigma^{X,head}$ are the ``body'' and ``head'' components of the exchange self-energy, respectively. The former body matrix element is expressed as 
\begin{eqnarray} 
\Sigma^{X,body}_{\alpha\beta{\bf k}} &=& \frac{4\pi}{V}\sum_{{\bf qG}n}{}^{\prime} M_{\alpha n}^{{\bf G}}({\bf k,\!k\!-\!q}) M_{\beta n}^{{\bf G}}({\bf k,\!k\!-\!q})^* \nonumber \\
&\times& \frac{1-\cos(|{\bf q\!+\!G}|R_c)}{|{\bf q\!+\!G}|^2} \tilde{\theta}(\mu-\epsilon_{n{\bf k-q}}) \label{Sx} 
\end{eqnarray} 
with 
\begin{eqnarray} 
\tilde{\theta}(\mu-\epsilon)=\frac{1}{\pi}\arctan \Bigl( \frac{\mu-\epsilon}{\delta}\Bigr)+\frac{1}{2}. \label{theta} 
\end{eqnarray} 
The prime in the sum of Eq.~(\ref{Sx}) represents the summation excluding the contribution of the head term of ${\bf q+G=0}$. Corresponding to the replacement of $v$ with $\tilde{v}$, the related Fourier transform is modified from $\frac{1}{|{\bf q+G}|}$ to $\frac{\sqrt{1-\cos(|{\bf q+G}|R_c)}}{|{\bf q+G}|}$ in Eq.~(\ref{Sx}). Accordingly, the body matrix element in Eq.~(\ref{Sx}) is supplemented by the head component in Eq.~(\ref{Sxb+Sxh})   
\begin{eqnarray}  
\Sigma^{X,head}_{\alpha\beta{\bf k}}=\frac{2\pi}{V} R_c^2 \delta_{\alpha\beta} \theta(\mu-\epsilon_{\alpha{\bf k}}).
\end{eqnarray}  

The operator of the correlation self-energy is defined by  
\begin{eqnarray}
\Sigma^C({\bf r,\!r'},\!\omega)\!=\!i\!\int\!\frac{d\omega'}{2\pi} G_0({\bf r,\!r'},\!\omega\!+\!\omega') W_C({\bf r,\!r'},\!\omega'),    
\label{GW_SIGMA}
\end{eqnarray}
where $W_C(\omega)=W(\omega)-v$ is the correlation part of the symmetrized screened Coulomb interaction $W(\omega)=v^{\frac{1}{2}}\epsilon^{-1}(\omega)v^{\frac{1}{2}}$. Here, $\epsilon^{-1}$ is the inverse dielectric function, which is calculated by inverting the symmetrized dielectric matrix in Eqs.~(\ref{symeps}) and (\ref{eq0}). For a practical calculation of the matrix element of $\Sigma^C({\bf r,r'},\omega)$ in Eq.~(\ref{GW_SIGMA}), we introduce the following model screened interaction~\cite{Nohara}  
\begin{eqnarray}
\tilde{W}_C({\bf r,r'},\omega_i)=\sum_{j} b_{ij}a_j({\bf r,r'})\label{modelW}
\end{eqnarray}
with
\begin{eqnarray}
b_{ij}=\frac{1}{\omega_i-z_j}-\frac{1}{\omega_i+z_j}. \label{pole-z}
\end{eqnarray}
In Eq.~(\ref{modelW}), the real frequency $\omega$ is discretized into $\omega_i$, and $z_j$ and $a_j({\bf r,r'})$ are the pole and amplitude of the model interactions, respectively. The matrix element $b_{ij}$ comprises a square matrix [see Sec.~III(A)]. Since the frequency-dependent part $\tilde{W}_C$ is decoupled from the amplitude one, the frequency integral in $iG_0\tilde{W}_C$ can be analytically performed. The matrix element of $\Sigma^C(\omega)$ consists of the body and head components as follows:  
\begin{eqnarray}
\Sigma^C_{\alpha\beta{\bf k}}(\omega)=\Sigma^{C,body}_{\alpha\beta{\bf k}}(\omega)+\Sigma^{C,head}_{\alpha\beta{\bf k}}(\omega). \label{Scb+Sch} 
\end{eqnarray}
The body matrix element in the above is given by  
\begin{eqnarray}
\Sigma^{C,body}_{\alpha\beta{\bf k}}(\omega)\!=\!\sum_{jn{\bf q}}{}^{\prime} 
\frac{\langle\phi_{\alpha{\bf k}}\phi_{n{\bf k-q}}|a_j|\phi_{n{\bf k-q}}\phi_{\beta{\bf k}}\rangle} 
{\omega\!-\!\epsilon_{n{\bf k-q}}\!-\!(z_j\!-\!i\delta){\rm sgn}(\epsilon_{n{\bf k-q}}\!-\!\mu)}, \label{eq:SGM} \nonumber \\ 
\end{eqnarray}
where the numerator is given by 
\begin{eqnarray}
\langle&\phi_{\alpha{\bf k}}&\phi_{n{\bf k-q}}|a_j|\phi_{n{\bf k-q}}\phi_{\beta{\bf k}}\rangle \nonumber \\ 
&=& \sum_{i} \bigl({\bf b}^{-1}\bigr)_{ji} 
\langle\phi_{\alpha{\bf k}}\phi_{n{\bf k-q}}|W_C(\omega_i)|\phi_{n{\bf k-q}}\phi_{\beta{\bf k}}\rangle
\end{eqnarray}
with 
\begin{eqnarray}
\langle&\phi_{\alpha{\bf k}}&\phi_{n{\bf k-q}}|W_C(\omega_i)|\phi_{n{\bf k-q}}\phi_{\beta{\bf k}}\rangle \nonumber \\ 
&=& \frac{4\pi}{V} \sum_{{\bf GG'}}{}^{\prime} \frac{M_{\alpha n}^{{\bf G}}({\bf k,k-q})\sqrt{1-\cos(|{\bf q+G}|R_c)}}{|{\bf q+G}|} \nonumber \\ 
&\times& \Bigl(\epsilon^{-1}_{{\bf GG'}}({\bf q},\omega_i)-\delta_{{\bf GG'}}\Bigr) \nonumber \\ 
&\times& \frac{M_{\beta n}^{{\bf G'}}({\bf k,k-q})^*\sqrt{1-\cos(|{\bf q+G'}|R_c)}}{|{\bf q+G'}|}. \label{Wc} 
\end{eqnarray} 
Note that, in the practical calculation, the frequency $\omega$ for the self-energy in Eq.~(\ref{eq:SGM}) is distinguished from the frequency $\omega_i$ for the screened interaction in Eq.~(\ref{modelW}). The body matrix element $\Sigma^{C,body}_{\alpha\beta{\bf k}}(\omega)$ in Eq.~(\ref{eq:SGM}) is supplemented with the head component in Eq.~(\ref{Scb+Sch})    
\begin{eqnarray}
\Sigma^{C,head}_{\alpha\beta{\bf k}}(\omega)\!=\!\frac{2\pi R_c^2}{V}\delta_{\alpha\beta}\sum_{j} \frac{g_j}{\omega\!-\!\epsilon_{\alpha{\bf k}}\!-\!(z_j\!-\!i\delta){\rm sgn}(\epsilon_{\alpha{\bf k}}\!-\!\mu)} \nonumber \\ 
\end{eqnarray} 
with 
\begin{eqnarray}
g_j=\sum_{i} \bigl({\bf b}^{-1}\bigr)_{ji} \Bigl( \epsilon_{\bf 00}^{-1}({\bf 0},\omega_i)-1 \Bigr). 
\end{eqnarray} 

With these ingredients, the spectral function is calculated via the Wannier-interpolation method (see below). The spectral function at an arbitrary ${\bf k}$ is \begin{eqnarray} 
A({\bf k},\omega)=\frac{1}{\pi} \sum_{\alpha} \Bigl| {\rm Im} \frac{1}{\omega-({\cal E}_{\alpha {\bf k}}(\omega)+\Delta)} \Bigr|, \label{Akw}
\end{eqnarray}
where ${\cal E}_{\alpha{\bf k}}(\omega)$ is obtained by diagonalizing non-symmetric complex matrix in the Wannier basis 
\begin{eqnarray} 
{\cal H}_{ij}({\bf k},\omega)=h_{ij}({\bf k})+\Sigma_{ij}({\bf k},\omega), \label{hij}
\end{eqnarray}
where $h_{ij}({\bf k})$ is the Fourier transform of the KS Hamiltonian matrix in the Wannier basis as   
\begin{eqnarray} 
h_{ij}({\bf k})=\sum_{{\bf R}} h_{ij{\bf R}} e^{i{\bf kR}} 
\end{eqnarray}
with  
\begin{eqnarray}
h_{ij{\bf R}} = \frac{1}{N} \sum_{{\bf k'}\alpha} \langle w_{i{\bf 0}}|\phi_{\alpha{\bf k'}} \rangle \epsilon_{\alpha{\bf k'}} \langle \phi_{\alpha{\bf k'}}|w_{j{\bf 0}} \rangle e^{i{\bf k'R}}. 
\end{eqnarray} 
Here, ${\bf k'}$ is a $k$ point in the regular mesh and $N$ is the total number of the $k$ points in the regular mesh. Also, $|w_{i{\bf R}}\rangle$ is the $i$th Wannier orbital at the lattice point ${\bf R}$, and the transform $\langle\phi_{\alpha{\bf k'}}|w_{i{\bf 0}}\rangle$ is obtained in the Wannier-function-generation routine. $\Sigma_{ij}({\bf k},\omega)$ in Eq.~(\ref{hij}) is the Fourier transform of the self-energy in the Wannier basis as 
\begin{eqnarray}
\Sigma_{ij}({\bf k},\omega)=\sum_{{\bf R}} \Sigma_{ij{\bf R}}(\omega) e^{i{\bf kR}}  \label{Sijkw} 
\end{eqnarray} 
with 
\begin{eqnarray}
\Sigma_{ij{\bf R}}(\omega)\!=\!\frac{1}{N}\!\!\sum_{{\bf k'}\alpha\beta}\!\langle w_{i{\bf 0}}|\phi_{\alpha{\bf k'}}\rangle\!\Sigma_{\alpha\beta{\bf k'}}(\omega)\!\langle \phi_{\beta{\bf k'}}|w_{j{\bf 0}}\rangle e^{i{\bf k'R}}.  
\nonumber \\ \label{SijRw} 
\end{eqnarray}
The matrix element $\Sigma_{\alpha\beta{\bf k'}}(\omega)$ is defined by 
\begin{eqnarray}
\Sigma_{\alpha\beta{\bf k'}}(\omega)=\langle\phi_{\alpha{\bf k'}}|\Sigma^{X}+\Sigma^{C}(\omega)-V^{xc}|\phi_{\beta{\bf k'}}\rangle  \label{Sigma-od} 
\end{eqnarray} 
with $V^{xc}$ being the exchange-correlation potential in density-functional theory. 

In Eq.~(\ref{Akw}), the energy shift $\Delta$ is introduced to correct the mismatch of the Fermi level between the KS and one-shot $GW$ systems. This parameter is determined from the equation on the spectral norm 
\begin{eqnarray}
\frac{2}{N_{{\bf k}}}\sum_{{\bf k}} \int_{-\infty}^{\mu} A({\bf k},\omega)d\omega=N_{{\rm elec}}, \label{Norm} 
\end{eqnarray} 
where $N_{{\rm elec}}$ is the total number of electrons in the system and $N_{{\bf k}}$ is the total number of sampling $k$ points after the interpolation. Note that $\mu$ is set to the Fermi level for the KS system.  

The flow of the calculation is as follows: We first perform density-functional calculations to obtain the band structures and the Wannier functions for bands associated with the self-energy calculations. We then calculate the self-energies for the irreducible $k$-points $\{{\bf \bar{k}}\}$ in a regular mesh, including band off-diagonal terms as $\Sigma_{\alpha\beta{\bf \bar{k}}}(\omega)=\Sigma^{X}_{\alpha\beta{\bf \bar{k}}}+\Sigma^{C}_{\alpha\beta{\bf \bar{k}}}(\omega)-V^{xc}_{\alpha\beta{\bf \bar{k}}}$ in the selected energy region. The self-energies at a $k'$ point symmetrically equivalent to ${\bf \bar{k}}$ are the same as $\Sigma_{\alpha\beta{\bf \bar{k}}}(\omega)$, but, in the time-reversal symmetry case, $\Sigma_{\alpha\beta{\bf k'}}(\omega)$ is obtained by $\Sigma_{\alpha\beta{\bf k'}}(\omega)=\Sigma_{\beta\alpha,{\bf -\bar{k}}}(\omega)$. Then, we transform $\Sigma_{\alpha\beta{\bf k'}}(\omega)$ in Eq.~(\ref{Sigma-od}) to the Wannier representation $\Sigma_{ij{\bf R}}(\omega)$ with Eq.~(\ref{SijRw}). With these data, we evaluate the self-energy at an arbitrary $\bf{ k}$ via Eq.~(\ref{Sijkw}). Note that the calculated spectral function includes band-off-diagonal effects, which is discussed in Appendix A. Finally, we calculate the spectral function of Eq.~(\ref{Akw}) considering the energy shift $\Delta$ in Eqs.~(\ref{Akw}) and (\ref{Norm}).  

\subsection{GW+cumulant expansion method} 
The $GW+C$ approach is a theory beyond the $GW$ approximation,~\cite{Aryasetiawan,GWC-1,GWC-2} which is based on systematic diagrammatic expansions. This approach is suitable for dealing with long-range correlations, i.e., various types of the plasmon-fluctuation diagrams not included in the usual $GW$ diagram. In the initial stage of the study, it was applied to a system of core electrons interacting with a plasmon field.~\cite{GWC-1} Currently, {\em ab initio} $GW$+$C$ calculations have been known to give a better description for satellites due to the plasmon excitation.~\cite{Aryasetiawan,Gatti,GWC-1,GWC-2,GWC-3,GWC-4,GWC-5,GWC-6,GWC-7}  

The Green's function with the cumulant expansion is defined in the time domain by~\cite{Aryasetiawan}   
\begin{eqnarray}
G_{\alpha{\rm k}}(t)=i \Theta(-t) e^{-i\epsilon_{\alpha{\bf k}}t+C_{\alpha{\bf k}}^h(t)}-i \Theta(t) e^{-i\epsilon_{\alpha{\bf k}}t+C_{\alpha{\bf k}}^p(t)}. \nonumber \\
\end{eqnarray}
where $\epsilon_{\alpha{\bf k}}<\mu$ for the first term on the right hand side and $\epsilon_{\alpha{\bf k}}>\mu$ for the second term. The $C_{\alpha{\bf k}}^h(t)$ and $C_{\alpha{\bf k}}^p(t)$ are the cumulants for the hole and particle states, respectively. The spectral function is calculated by the Fourier transform as 
\begin{eqnarray}
A({\bf k},\omega) 
 &=& \frac{1}{\pi}\sum_{\alpha}{\rm Im} \int_{-\infty}^{\infty} dt 
     e^{i\omega t} G_{\alpha{\bf k}}(t) \nonumber \\   
 &=& A^h({\bf k},\omega)+A^p({\bf k},\omega), \label{AkwGWC} 
\end{eqnarray}
which consists of the hole $A^h({\bf k},\omega)$ and particle $A^p({\bf k},\omega)$ contributions. 

The spectral function for the hole part is written as 
\begin{eqnarray}
A^h({\bf k},\omega)\!=\!\frac{1}{\pi}\!\sum_{\alpha}^{{\rm occ}}{\rm Im}i\!\int_{-\infty}^{0}\!dt e^{i\omega t} e^{-i \epsilon_{\alpha{\bf k}}t}  
e^{C_{\alpha{\bf k}}^{h}(t)}\!, \label{AhkwGWC}   
\end{eqnarray}
where the band sum is taken over the occupied states. To the lowest order in the screened interaction $W$, the cumulant is obtained by~\cite{Aryasetiawan,GWC-1,GWC-2,GWC-note} 
\begin{eqnarray} 
C_{\alpha{\bf k}}^h(t)=i\int_{t}^{\infty}dt'\int_{t'}^{\infty}d\tau e^{i\epsilon_{\alpha{\bf k}} \tau}\Sigma_{\alpha{\bf k}}(\tau),  \label{Chdef} 
\end{eqnarray}
where $\Sigma=\Sigma^{X}+\Sigma^{C}-V^{xc}$.  

In the present study, the cumulant is expanded around the quasiparticle energy $E_{\alpha{\bf k}}$,~\cite{GWC-5} which is a solution of 
\begin{eqnarray}
E_{\alpha{\bf k}}=\epsilon_{\alpha{\bf k}}+{\rm Re}\Sigma_{\alpha{\bf k}}(E_{\alpha{\bf k}}) +\Delta,      
\end{eqnarray} 
where $\Delta$ is an energy shift to correct a mismatch of the Fermi level between the KS and one-shot $GW$+$C$ systems.  
By considering the Fourier transform of $\Sigma(\tau)$ in Eq.~(\ref{Chdef}) and the spectral representation of $\Sigma(\omega)$, and after some manipulation,~\cite{Aryasetiawan,GWC-1,GWC-2} the expression of the cumulant is obtained, which consists of the quasiparticle and satellite parts as 
\begin{eqnarray} 
C_{\alpha{\bf k}}^{h}(E_{\alpha{\bf k}},t)=C_{\alpha{\bf k}}^{h,qp}(E_{\alpha{\bf k}},t)+C_{\alpha{\bf k}}^{h,s}(E_{\alpha{\bf k}},t) \label{Ctoth} 
\end{eqnarray} 
with 
\begin{eqnarray}
C_{\alpha{\bf k}}^{h,qp}(E_{\alpha{\bf k}},t)\!=\!-i(\Sigma_{\alpha{\bf k}}(E_{\alpha{\bf k}})\!+\!\Delta)t\!+\!\frac{\partial \Sigma_{\alpha{\bf k}}^h(E_{\alpha{\bf k}})}{\partial \omega} \label{Cqph}
\end{eqnarray}
and  
\begin{eqnarray}
C_{\alpha{\bf k}}^{h,s}(E_{\alpha{\bf k}},t)=\frac{1}{\pi} \int_{-\infty}^{\mu}  d\omega' \frac{e^{i(E_{\alpha{\bf k}}-\omega'-i\delta)t}}{(E_{\alpha{\bf k}}-\omega'-i\delta)^2} {\rm Im} \Sigma_{\alpha{\bf k}}(\omega'). \nonumber \\ \label{Csh} 
\end{eqnarray}
Note that $t$ is negative for the hole part. To show the expansion point explicitly, we add $E_{\alpha{\bf k}}$ as an index in the cumulant Eqs.~(\ref{Ctoth}), (\ref{Cqph}), and (\ref{Csh}). Within the one-shot calculation, the position of the cumulant expansion may be taken at the non-interacting energy $\epsilon_{\alpha{\bf k}}$.~\cite{Aryasetiawan,GWC-1,GWC-2} By taking the expansion point at $E_{\alpha{\bf k}}$, the results may include some sort of the self-consistency effect. 

The hole self-energy in Eq.~(\ref{Cqph}) is defined by 
\begin{eqnarray}
\Sigma_{\alpha{\bf k}}^h(\omega)=\frac{1}{\pi}\int_{\infty}^{\mu} d\omega' \frac{{\rm Im}\Sigma_{\alpha{\bf k}}(\omega')}{\omega-\omega'-i\delta}. 
\end{eqnarray}
It should be noted that the derivative $\frac{\partial\Sigma_{\alpha{\bf k}}^h(E_{\alpha{\bf k}})}{\partial\omega}$ in Eq.~(\ref{Cqph}) is related to the $t$=0 component of the satellite cumulant in Eq.~(\ref{Csh}) as~\cite{Aryasetiawan}
\begin{eqnarray}
\frac{\partial\Sigma_{\alpha{\bf k}}^h(E_{\alpha{\bf k}})}{\partial\omega} 
&=& -C_{\alpha{\bf k}}^{h,s}(E_{\alpha{\bf k}},0) \nonumber \\ 
&=& -\frac{1}{\pi} \int_{-\infty}^{\mu} d\omega' \frac{{\rm Im}\Sigma_{\alpha{\bf k}}(\omega')}{(E_{\alpha{\bf k}}-\omega'-i\delta)^2}. \label{del-eq-C}
\end{eqnarray} 
This expression is practically used for the evaluation of the derivative; we avoid the numerical calculation of the derivative by a finite difference and use the integral expression on the right hand side of Eq.~(\ref{del-eq-C}) for the derivative, since we find that the latter treatment is numerically more stable. The stable calculation of the derivative is important in keeping the sum rule on the $GW+C$ spectrum. 

In the practical calculation, we divide $A^{h}({\bf k},\omega)$ in Eq.~(\ref{AhkwGWC}) into the two parts for stable calculations as~\cite{Aryasetiawan}  
\begin{eqnarray}
A^{h}({\bf k},\omega)\!=\!A^{h,qp}({\bf k},\omega)\!+\!A^{h,qp}({\bf k},\omega)\!\ast\!A^{h,s}({\bf k},\omega). \label{ahkw} 
\end{eqnarray} 
The first quasiparticle term on the right hand side can be calculated analytically as  
\begin{eqnarray}
A^{h,qp}({\bf k},\omega)\!=\!\frac{1}{\pi}\sum_{\alpha}^{{\rm occ}}e^{-\gamma_{\alpha{\bf k}}} \frac{\eta_{\alpha{\bf k}}\cos\beta_{\alpha{\bf k}}-(\omega-E_{\alpha{\bf k}})\sin\beta_{\alpha{\bf k}} }{(\omega-E_{\alpha{\bf k}})^2+\eta_{\alpha{\bf k}}^2},  \nonumber \\
\end{eqnarray}
where 
\begin{eqnarray}
\eta_{\alpha{\bf k}}&=&{\rm Im}\Sigma_{\alpha{\bf k}}(E_{\alpha{\bf k}})+\delta, \label{ImSGMh} \\
\gamma_{\alpha{\bf k}}&=&-{\rm Re}\frac{\partial\Sigma_{\alpha{\bf k}}^h(E_{\alpha{\bf k}})}{\partial \omega}, \\
\beta_{\alpha{\bf k}}&=&-{\rm Im}\frac{\partial\Sigma_{\alpha{\bf k}}^h(E_{\alpha{\bf k}})}{\partial \omega}. 
\end{eqnarray} 
The latter convolution term in Eq.~(\ref{ahkw}) is calculated via the numerical integration as 
\begin{widetext} 
\begin{eqnarray} 
A^{h,qp}({\bf k},\omega)\ast A^{h,s}({\bf k},\omega) =\frac{1}{\pi} \sum_{\alpha}^{{\rm occ}} {\rm Im} i \int_{-\infty}^{0} dt e^{i\omega t} e^{-i \epsilon_{\alpha{\bf k}t}}  e^{C_{\alpha{\bf k}}^{h,qp}(E_{\alpha{\bf k}},t)} \bigl(e^{C_{\alpha{\bf k}}^{h,s}(E_{\alpha{\bf k}},t)}-1\bigr),   \label{Ash}  
\end{eqnarray}
\end{widetext} 
where the integrand in the right-hand side decays to zero rapidly. 

The particle part of the spectral function in Eq.~(\ref{AkwGWC}) is given by  
\begin{eqnarray}
A^p({\bf k},\omega)\!=\!\frac{1}{\pi}\!\sum_{\alpha}^{{\rm unocc}}{\rm Im}i\!\int_{0}^{\infty}\!dt e^{i\omega t} e^{-i \epsilon_{\alpha{\bf k}t}}  
e^{C_{\alpha{\bf k}}^{p}(E_{\alpha{\bf k}},t)}  \label{ApkwGWC} 
\end{eqnarray}
with 
\begin{eqnarray} 
C_{\alpha{\bf k}}^{p}(E_{\alpha{\bf k}},t)=C_{\alpha{\bf k}}^{p,qp}(E_{\alpha{\bf k}},t)+C_{\alpha{\bf k}}^{p,s}(E_{\alpha{\bf k}},t). 
\end{eqnarray} 
In this case $t$ is positive. The band sum in Eq.~(\ref{ApkwGWC}) runs over the unoccupied states. The $C_{\alpha{\bf k}}^{p,qp}(E_{\alpha{\bf k}},t)$ and $C_{\alpha{\bf k}}^{p,s}(E_{\alpha{\bf k}},t)$ are given by  
\begin{eqnarray}
C_{\alpha{\bf k}}^{p,qp}(E_{\alpha{\bf k}},t)\!=\!-i(\Sigma_{\alpha{\bf k}}(E_{\alpha{\bf k}})\!+\!\Delta)t\!+\!\frac{\partial\Sigma_{\alpha{\bf k}}^p(E_{\alpha{\bf k}})}{\partial\omega}\! \label{Cqpp} 
\end{eqnarray}
and 
\begin{eqnarray}
C_{\alpha{\bf k}}^{p,s}(E_{\alpha{\bf k}},t)=\frac{-1}{\pi} \int_{\mu}^{\infty} d\omega' \frac{e^{i(E_{\alpha{\bf k}}-\omega'+i\delta)t}}{(E_{\alpha{\bf k}}-\omega'+i\delta)^2} {\rm Im} \Sigma_{\alpha{\bf k}}(\omega'), \label{Csp} \nonumber \\ 
\end{eqnarray}
respectively. 
The $\Sigma^p$ in Eq.~(\ref{Cqpp}) is the particle self-energy as  
\begin{eqnarray}
\Sigma_{\alpha{\bf k}}^p(\omega)=\frac{-1}{\pi}\int_{\mu}^{\infty} d\omega' \frac{{\rm Im}\Sigma_{\alpha{\bf k}}(\omega')}{\omega-\omega'+i\delta}         
\end{eqnarray}
 Note that the self-energies in the above equations are defined as the causal one, so that the imaginary part of the self-energy is positive for $\omega<\mu$ and negative for $\omega>\mu$. In the particle part, $\frac{\partial\Sigma_{\alpha{\bf k}}^p(E_{\alpha{\bf k}})}{\partial\omega}=-C_{\alpha{\bf k}}^{p,s}(E_{\alpha{\bf k}},0)$  holds similarly to Eq.~(\ref{del-eq-C}). The contribution from the quasiparticle part to $A^{p}({\bf k},\omega)$ is 
\begin{eqnarray}
A^{p,qp}({\bf k},\omega)=\frac{-1}{\pi}\sum_{\alpha}^{{\rm unocc}} e^{-\gamma_{\alpha{\bf k}}} \frac{\eta_{\alpha{\bf k}}\cos\beta_{\alpha{\bf k}}-(\omega-E_{\alpha{\bf k}})\sin\beta_{\alpha{\bf k}} }{(\omega-E_{\alpha{\bf k}})^2+\eta_{\alpha{\bf k}}^2} \nonumber \\
\end{eqnarray}
with  
\begin{eqnarray}
\eta_{\alpha{\bf k}}&=&{\rm Im}\Sigma_{\alpha{\bf k}}(E_{\alpha{\bf k}})-\delta, \label{ImSGMp} \\
\gamma_{\alpha{\bf k}}&=&-{\rm Re}\frac{\partial\Sigma_{\alpha{\bf k}}^p(E_{\alpha{\bf k}})}{\partial \omega}, \\
\beta_{\alpha{\bf k}}&=&-{\rm Im}\frac{\partial\Sigma_{\alpha{\bf k}}^p(E_{\alpha{\bf k}})}{\partial\omega},
\end{eqnarray}
 and the convolution-part contribution is 
\begin{widetext} 
\begin{eqnarray} 
A^{p,qp}({\bf k},\omega)\ast A^{p,s}({\bf k},\omega) =\frac{1}{\pi} \sum_{\alpha}^{{\rm unocc}}{\rm Im} i \int_{0}^{\infty} dt e^{i\omega t} e^{-i \epsilon_{\alpha{\bf k}t}}  e^{C_{\alpha{\bf k}}^{p,qp}(E_{\alpha{\bf k}},t)} \bigl(e^{C_{\alpha{\bf k}}^{p,s}(E_{\alpha{\bf k}},t)}-1\bigr). \label{Asp} 
\end{eqnarray}
\end{widetext}

\section{results and discussions}\label{sec:result}
\subsection{Calculation condition}
Density-functional calculations were performed with {\sf{Tokyo Ab-initio Program Package}}~\cite{TAPP} with plane-wave basis sets, where we employed norm-conserving pseudopotentials~\cite{PP-1,PP-2} and generalized gradient approximation (GGA) for the exchange-correlation potential.~\cite{PBE96} Maximally localized Wannier functions~\cite{MaxLoc-1,MaxLoc-2} were used for the interpolation of the self-energy. 

For the atomic coordinates of (TMTSF)$_2$PF$_6$, the experimental structure obtained by a neutron measurement~\cite{structure-TMTSF} at 20 K was adopted. The cutoff energies in wavefunction and in charge densities are 36 Ry and 144 Ry, respectively, and a 15$\times$15$\times$3 $k$-point sampling was employed. The cutoff for the polarization function in Eq.~(\ref{eq:chi}) was set to be 3 Ry, and 200 bands were considered, which covers an energy range from the bottom of the occupied states near $-$30 eV to the top of the unoccupied states near 15 eV, where 0 eV is the Fermi level. The frequency grid of the polarization was taken up to $\omega_{max}$ = 86 eV in a double-logarithmic form, for which we sampled 99 energy-points for [0.1 eV: 43 eV] and 10 points for [43 eV: 86 eV] with initial grid set to be 0.0 eV. The $k$ sum over BZ in Eqs.~(\ref{eq:chi}), (\ref{eq0}), and (\ref{wpl}) was evaluated by the generalized tetrahedron method.~\cite{Fujiwara,Nohara} The Drude term of $\omega=0$ is evaluated at the slightly shifted frequency $\omega=10^{-10}$ (au). The self-energy in Eq.~(\ref{eq:SGM}) was calculated for the frequency range [$-\Omega_{max}$ eV: $\Omega_{max}$ eV], where $\Omega_{max}$ was set to be 100 eV. In the practical calculation, we sampled 450 points for [$-$100 eV: $-$10 eV] with the interval of 0.2 eV, 1000 points for [$-$10 eV: 10 eV] with the 0.02 eV interval, and 450 points for [10 eV: 100 eV] with the 0.2 eV interval. With this energy range, the high-frequency tail of the self-energy is sufficiently small. This convergence is important to preserve the norm of the spectral function. 

For SrVO$_3$, band calculations were performed for the idealized simple cubic structure, where the lattice parameter was set to be $a$=3.84 {\AA}. The cutoff energies for wavefunction and charge densities are 49 Ry and 196 Ry, respectively, and an 11$\times$11$\times$11 $k$-point sampling was employed. The cutoff energy for the polarization function was set to be 10 Ry and 130 bands were considered, which cover from the bottom of the occupied states near $-$20 eV to the top of the unoccupied states near 90 eV. The frequency range of the polarization function was taken to be $\omega_{max}=220$ eV, where finer logarithmic sampling was applied to [0.1 eV: 110 eV] with 189 points and coarser sampling was done for [110 eV: 220 eV] with 10 points, with initial grid set to be 0.0 eV. In the self-energy calculation, $\Omega_{max}$ was set to be 200 eV, thus, the frequency dependence of the self-energy was calculated for [$-$200 eV: 200 eV], where we sampled 200 points for [$-$200 eV: $-$40 eV] with the 0.8 eV interval, 1600 points for [$-$40 eV: 40 eV] with the 0.05 eV interval, and 200 points for [40 eV: 200 eV] with the 0.8 eV interval.  

In the fitting of the model screened interaction [Eqs.~(\ref{modelW}) and (\ref{pole-z})], the positions of the poles in the model interaction are set as follows:~\cite{Nohara}     
\begin{eqnarray}
z_i=\frac{\omega_{i+1}+\omega_i}{2}+i\Bigl(\frac{3}{2}\Delta_i\Bigr) 
\end{eqnarray} 
with $\Delta_i=\omega_{i+1}-\omega_i$. The total number of \{$z_i$\} is the same as that of the frequency grid \{$\omega_i$\} for the polarization function. We checked that the {\em ab initio} screened interaction $W_C(\omega)$ are satisfactorily reproduced by this model function $\tilde{W}_C(\omega)$. The broadening $\delta$ in Eqs.~(\ref{eq:chi}), (\ref{theta}), (\ref{eq:SGM}), (\ref{Csh}), (\ref{ImSGMh}), (\ref{Csp}), and (\ref{ImSGMp}) was set to be 0.02 eV for (TMTSF)$_2$PF$_6$ and 0.05 eV for SrVO$_3$. The value of $\delta$ is desirable to be small enough, but the lower-bound in the practical calculation is determined by the resolution of the band dispersion, depending primary on the $k$-mesh density. Also, the cutoff $R_c$ in the attenuation potential [Eqs.~(\ref{Sx}) and (\ref{Wc})] is 28.27 {\AA} for (TMTSF)$_2$PF$_6$ and 11.53 {\AA} for SrVO$_3$, respectively. For (TMTSF)$_2$PF$_6$, the shift $\Delta$ in the spectral function $A({\bf k}, \omega)$ in Eq.~(\ref{Akw}) was found to be 1.06 eV for the $GW$ calculation and 0.90 eV for the $GW$+$C$ one, respectively. For SrVO$_3$, $\Delta$ = 2.1 eV for the $GW$ and 2.35 eV for the $GW$+$C$. 

In $GW$+$C$, two numerical integrals on time and frequency appear [Eqs.~(\ref{Ash}) and (\ref{Asp}) for the time integral and Eqs.~(\ref{Csh}) and (\ref{Csp}) for frequency integral], which must be treated carefully. We performed time integrals in Eq.~(\ref{Ash}) numerically for the range [$-t_{max}$ au: 0 au] and those in Eq.~(\ref{Asp}) for [0 au: $t_{max}$ au], where $t_{max}$ is 50 (au). The total number of the time grid $N_t$ is 50000, with the interval $\Delta t=t_{max}\big/N_t=0.001$ au. Note that $\Omega_{max}\Delta t \ll 1$ is necessary to reproduce the norm of the spectral weight correctly. The frequency integral in Eqs.~(\ref{Csh}) and (\ref{Csp}) was numerically evaluated with the Simpson's formula for the interval $\Delta\omega$ divided into 21 subintervals, which is also important in obtaining the correct time dependence of the satellite cumulant. Also, in the $k$ integration to obtain the $GW+C$ density of states $A(\omega)$, the random $k$-point sampling was performed to improve the statistical average, where the Wannier interpolation technique was efficiently applied. With this condition, we obtained well converged spectra.           

\subsection{Density-functional band structure} 
\begin{figure}[htbp]
\vspace{0cm}
\begin{center}
\includegraphics[width=0.35\textwidth]{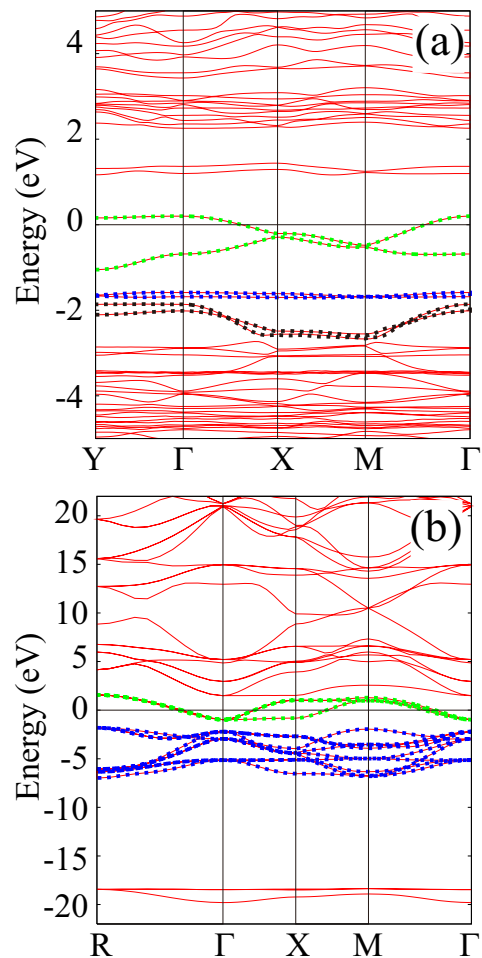}
\caption{(Color online) Density-functional GGA band structures (solid red curves) of (a) (TMTSF)$_2$PF$_6$ and (b) SrVO$_3$. The Fermi level is at zero energy. In the panel~(a), the green-dotted, blue-dotted, and black-dotted curves denote the HOMO, HOMO$-1$, and HOMO$-2$ bands, respectively. In (b), the green-dotted and blue-dotted curves correspond to the $t_{2g}$ and O$_{p}$ bands, respectively.}
\label{GGAband}
\end{center}
\end{figure}

Figure~\ref{GGAband} shows calculated GGA band structures of (TMTSF)$_2$PF$_6$ [panel (a)] and SrVO$_3$ [(b)]. In both systems, isolated bands are found around the Fermi level. In (TMTSF)$_2$PF$_6$, the isolated bands consist of highest-occupied molecular orbital (HOMO) of two molecules in the unit cell. [For detailed atomic geometry of (TMTSF)$_2$PF$_6$, refer to Refs.~\onlinecite{Kuroki} and \onlinecite{TMTSF-plasmon-Nakamura}.] We call them the ``HOMO" bands which are shown with the green-dotted curves. In addition, in this figure, ``HOMO$-$1" and ``HOMO$-$2" bands are shown by blue- and black-dotted curves, respectively.~\cite{Wannier-Orbital-TMTSF} In SrVO$_3$, the isolated bands (green-dotted curves) around the Fermi level are formed by the $t_{2g}$ orbitals of the vanadium atom and bands around [$-$7 eV: $-$2 eV] come from the oxygen-$p$ orbitals (blue-dotted curves).~\cite{Wannier-Orbital-SVO} 

\subsection{Low-energy plasmon excitation} 
\begin{figure}[htbp]
\vspace{0cm}
\begin{center}
\includegraphics[width=0.4\textwidth]{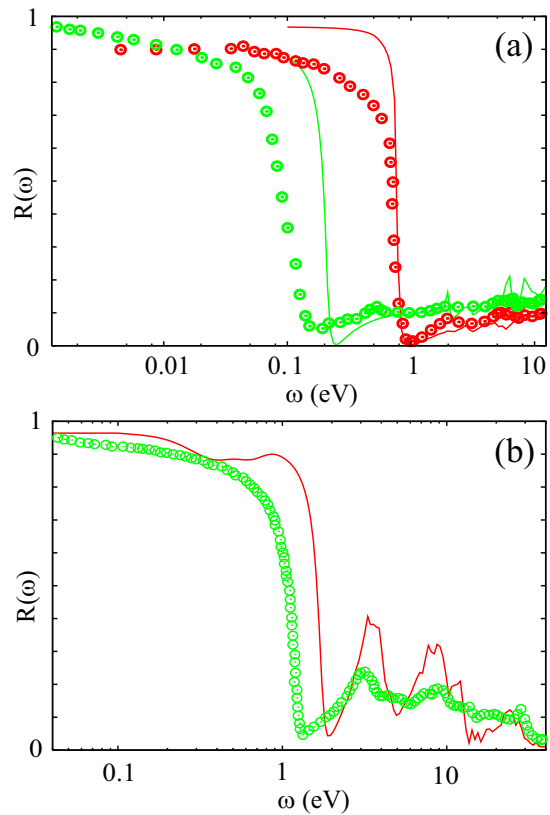}
\caption{(Color online) {\it Ab initio} reflectivity $R(\omega)$ (solid curve) based on the random-phase approximation and experimental one (open circles) of (a) (TMTSF)$_2$PF$_6$ measured at 25 K and (b) SrVO$_3$ measured at room temperature. The experimental data are taken from Ref.~\onlinecite{TMTSF-R-Dressel} for (TMTSF)$_2$PF$_6$ and Ref.~\onlinecite{R-SVO} for SrVO$_3$. In (TMTSF)$_2$PF$_6$, the results for $E\|a$ and $E\|b'$ are displayed by dark red and light green, respectively.}
\label{reflectance}
\end{center}
\end{figure}

To confirm the low-energy plasmon excitation in the above isolated-band systems, we calculated the reflectance spectra with the random-phase approximation  
\begin{eqnarray}
R_{\mu\mu}(\omega)=\Biggl| \frac{1-\sqrt{\epsilon_{\mu\mu}^{-1}(\omega)}}{1+\sqrt{\epsilon_{\mu\mu}^{-1}(\omega)}} \Biggr|,  
\end{eqnarray}
where $\epsilon_{\mu\mu}^{-1}(\omega)$ is obtained by inverse of the dielectric matrix in Eq.~(\ref{eq0}). Figure~\ref{reflectance} (a) is the result for (TMTSF)$_2$PF$_6$, where dark-red and light-green colors represent the results in the light polarization of $E\|a$ and $E\|b'$, respectively, and the $a$ axis ($a$$\perp$$b'$) is the one-dimensional conducting axis. The calculated results (solid curves) are compared with experimental results (circles). We see that the theoretical plasma edges satisfactorily reproduce the experimental ones around 0.8 eV for $E\|a$ and 0.1-0.2 eV for $E\|b'$, which indicates that the present scheme correctly captures the low-energy plasmon excitation. The panel (b) shows the result for SrVO$_3$. We again see a reasonable agreement between the theory and experiment for the plasma edge (1.8 eV for theory and 1.4 eV for the experiment).  

TABLE~\ref{PARAM-wWU} summarizes parameters characterizing low-energy electronic structures of the two isolated-band systems, i.e., the HOMO bands of (TMTSF)$_2$PF$_6$ and the $t_{2g}$ bands of SrVO$_3$. The table includes the calculated bare plasma frequency $\omega_{pl}$ in Eq.~(\ref{wpl}), bandwidth $W$, and effective local-interaction parameter $U-V$ with $U$ and $V$ being onsite and nearest-neighbor interactions, respectively, calculated with the constrained random phase approximation.~\cite{Nakamura-cRPA-1,Nakamura-cRPA-2,Nakamura-cRPA-3} We note that $\omega_{pl}$ is by definition different from the plasma edge in the reflectance spectra in Fig.~\ref{reflectance}; the latter energies are lowered from the bare $\omega_{pl}$ by the presence of the individual electronic excitations. We see that $\omega_{pl}$ has the same size as $W$ and $U-V$, indicating that energy scale of the plasmon excitation would compete with those of kinetic and local electronic-interaction energies of electrons in the isolated band. The long-range interaction and related plasmon fluctuations would clearly be important for electronic structure. 
 
\begin{table}[h] 
\caption{List of parameters for bare plasma frequency $\omega_{pl}$, bandwidth $W$, onsite interaction $U$, nearest-neighbor interaction $V$, and $U-V$ for (TMTSF)$_2$PF$_6$ and SrVO$_3$. The interaction parameters are calculated with the constrained random phase approximation.~\cite{Nakamura-cRPA-1,Nakamura-cRPA-2,Nakamura-cRPA-3} The parameters of (TMTSF)$_2$PF$_6$ are calculated for the HOMO bands, and those of SrVO$_3$ are evaluated for the $t_{2g}$ bands. The unit is eV.} 
\vspace{0.2cm} 
\centering 
\begin{tabular}{c@{\ \ \ \ }c@{\ \ \ \ }c@{\ \ \ \ }c@{\ \ \ \ }c@{\ \ \ \ }c} \hline \hline \\ [-8pt]
  & $\omega_{pl}$ & $W$ & $U$ & $V$ & $U-V$ \\ [3pt] \hline \\ [-5pt] 
\multirow{2}{*}{(TMTSF)$_2$PF$_6$} & 1.25 ($E\|a$) & \multirow{2}{*}{1.26} & 
\multirow{2}{*}{2.02} & \multirow{2}{*}{0.94} & \multirow{2}{*}{1.08} \\ 
                                   & 0.20 ($E\|b'$)& & & & \\ [4pt] 
SrVO$_3$ & 3.54 & 2.55 & 3.48 & 0.79 & 2.69 \\ [2pt] 
\hline \hline
\end{tabular} 
\label{PARAM-wWU} 
\end{table}

\subsection{Spectral function of (TMTSF)$_2$PF$_6$} 
\begin{figure}[htbp]
\vspace{0cm}
\begin{center}
\includegraphics[width=0.45\textwidth]{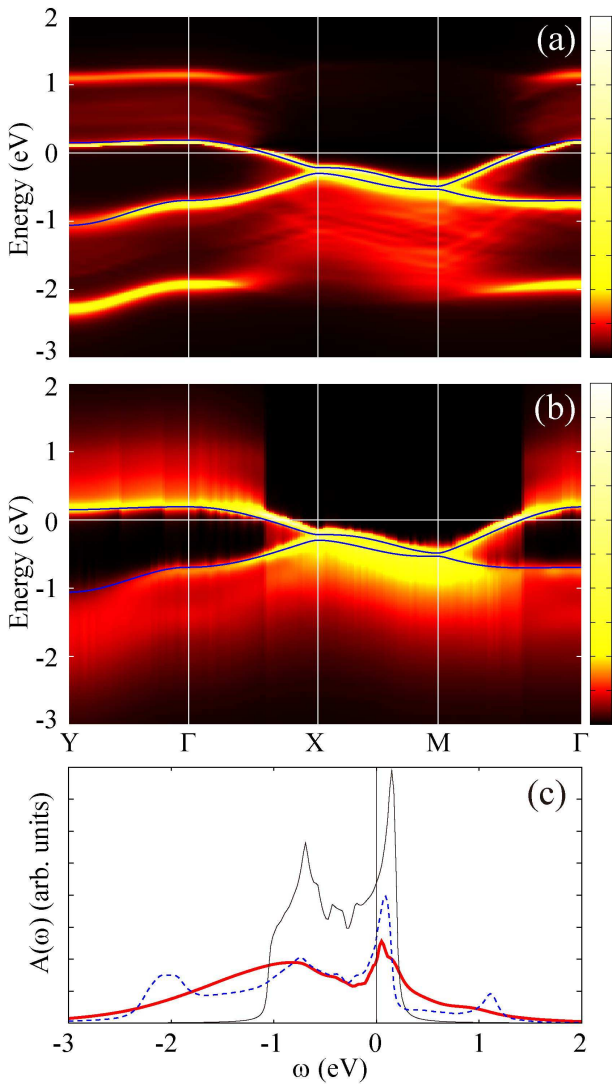}
\caption{(Color online) Spectral function for the HOMO bands of (TMTSF)$_2$PF$_6$ calculated with (a) $GW$ approximation and (b) $GW+C$ method. Blue-solid curves are the GGA results. The Fermi level is at zero energy. The colorbar is in linear scale. (c) Comparison of the density of states among the $GW+C$ (red-solid curve), $GW$ (blue-dotted curve), and GGA (black-thin curve) results. }
\label{Akw-TMTSF}
\end{center}
\end{figure}

To study effects of the low-energy plasmon excitation in the isolated bands on the electronic structure, we calculated a spectral function $A({\bf k},\omega)$ for the HOMO bands of (TMTSF)$_2$PF$_6$. Figure~\ref{Akw-TMTSF} displays the calculated spectra, where the panels (a) and (b) are the $GW$ result via Eq.~(\ref{Akw}) and the $GW$+$C$ one via Eq.~(\ref{AkwGWC}), respectively. For comparison, the GGA band structure is superposed with blue-solid curves. In the $GW$ spectrum, clear incoherent peaks appear; along the Y-$\Gamma$ line, the spectral intensities of plasmaron states~\cite{plasmaron-1,plasmaron-2} emerge about 1 eV above (below) the unoccupied (occupied) part of the HOMO bands.~\cite{TMTSF-plasmon-Nakamura} Also, along the X-M line, the spectra are more broadened and spread in the range from $-$1.5 to 0 eV. Interestingly, these sharp plasmaron peaks do not appear in the $GW$+$C$ spectrum. Instead, the $GW$+$C$ spectrum exhibits a broad incoherent structure throughout BZ. This is because the $GW$+$C$ treatment further takes into account the long-range correlation effect~\cite{Aryasetiawan,GWC-4} or various types of the self-energy diagram involving the plasmon fluctuation, which is not included in the standard $GW$ calculations.    

In the panel (c), density of states 
\begin{eqnarray}
A(\omega)=\int_{{\rm BZ}} A({\bf k},\omega) d{\bf k} 
\end{eqnarray} 
is shown, where the $GW$+$C$, $GW$, and GGA results are plotted by red-solid, blue-dotted, and thin-black-solid curves, respectively. Compared to the GGA spectrum, the $GW$ and $GW+C$ spectra show an appreciable band renormalization around the Fermi level by the plasmon excitation. We again confirm that the distinct plasmon satellite (plasmaron) around $-$2 eV and +1 eV in the $GW$ spectrum disappears in the $GW$+$C$ spectrum.  

\begin{figure}[htbp]
\vspace{0cm}
\begin{center}
\includegraphics[width=0.4\textwidth]{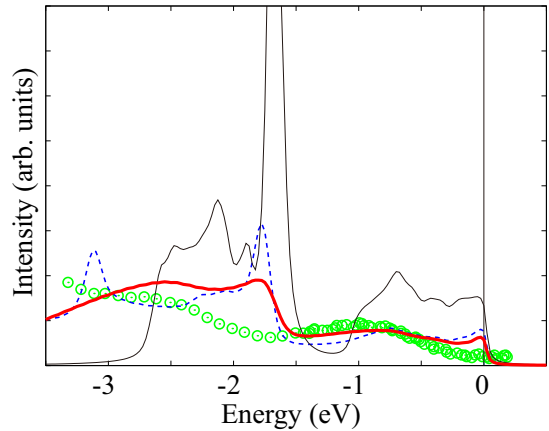}
\caption{(Color online) (a) {\it Ab initio} photo-emission spectra and experimental one measured with the HeII radiation at 50 K (green open circles) (Ref.~\onlinecite{TMTSF-PES-1}). The spectra are calculated for the HOMO, HOMO$-1$, and HOMO$-2$ bands. Red-thick-solid, blue-dotted, and black-thin-solid curves denote the $GW$+$C$, $GW$, and GGA results, respectively. A Lorentzian broadening of 0.02 eV is applied to the calculated spectra.}
\label{pes-TMTSF}
\end{center}
\end{figure}

Figure~\ref{pes-TMTSF} is the comparison between theoretical photo-emission spectra 
\begin{eqnarray}
A(\omega<\mu)=\int_{{\rm BZ}} A({\bf k},\omega<\mu) d{\bf k} 
\end{eqnarray} 
and experimental one (green open circles) obtained with HeII radiation ($h\nu$=40.8 eV) at 50 K (Ref.~\onlinecite{TMTSF-PES-1}). Thick-red-solid, blue-dotted, and thin-black-solid curves are the $GW$+$C$, $GW$, and GGA results, respectively. The spectra were calculated for the HOMO, HOMO$-1$, and HOMO$-2$ bands to cover the energy range measured in the experiment. We see that the GGA spectrum around the Fermi level is largely reduced in the $GW$ and $GW$+$C$ spectra by the self-energy effect due to the plasmon excitation. The $GW$+$C$ spectrum agrees with the overall profile of the experimental spectrum better than the $GW$ result; the broader spectrum is obtained in the $GW+C$ result around $-2\sim-$3 eV than in the $GW$ one. The discrepancy in the level position in this region between the theory and experiment arises probably from the level underestimation of the flat GGA HOMO$-$1 band [see Fig.~\ref{GGAband}~(a)]. 

\subsection{Spectral function of SrVO$_3$} 
\begin{figure}[htbp]
\vspace{0cm}
\begin{center}
\includegraphics[width=0.45\textwidth]{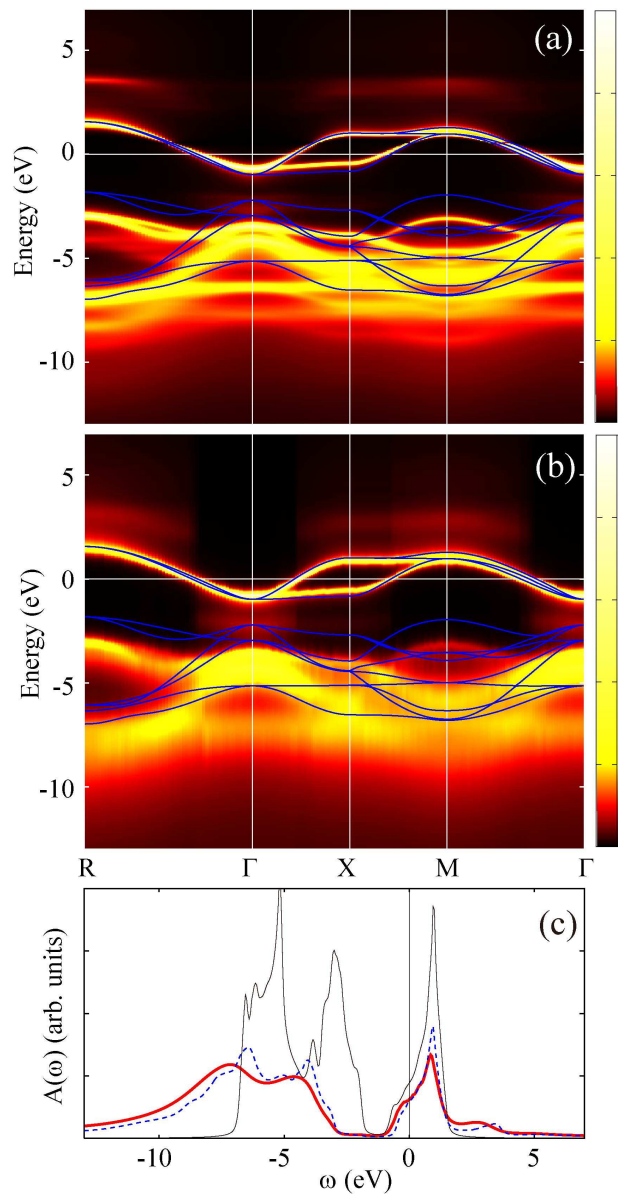}
\caption{(Color online) Spectral function for the $t_{2g}$ and O$_p$ bands of SrVO$_3$ calculated with (a) $GW$ approximation and (b) $GW+C$ method. Blue-solid curves are the GGA results. The Fermi level is at zero energy. The colorbar is in linear scale. (c) Comparison of the density of states among $GW$+$C$ (red-solid curve), $GW$ (blue-dotted curve), and GGA (black-thin curve) results. }
\label{Akw-SVO}
\end{center}
\end{figure}

Next, we consider the low-energy plasmon-fluctuation effect on the electronic structure of the transition-metal oxide SrVO$_3$. Figure~\ref{Akw-SVO} shows the spectral function calculated for the $t_{2g}$ and O$_p$ bands, where the panels (a) and (b) show the $GW$ and $GW$+$C$ results, respectively. The GGA bands are depicted with blue-solid curves. The low-energy plasmon satellite emerges around 1 eV above (below) the unoccupied (occupied) part of the $t_{2g}$ bands, but the intensity is weaker than that of (TMTSF)$_2$PF$_6$. Similarly to the TMTSF case, the $GW$+$C$ for SrVO$_3$ makes the plasmon satellite broader than the $GW$ result. On the O$_p$ bands, the self-energy effect is appreciable; the imaginary part of the self-energy, which is related to the lifetime of the quasiparticle states, becomes large as the binding energy is apart from the Fermi level. Thus, the spectrum of the O$_p$ band becomes rather broad. 

The panel (c) shows the comparison of calculated density of states. We see that the GGA spectrum in the $t_{2g}$ and O$_p$ bands is largely renormalized in the $GW$ and $GW+C$ spectra. The shape of the $GW$ spectrum resembles the $GW$+$C$ one, though the latter is somewhat more broadened.  

\begin{figure}[htbp]
\vspace{0cm}
\begin{center}
\includegraphics[width=0.4\textwidth]{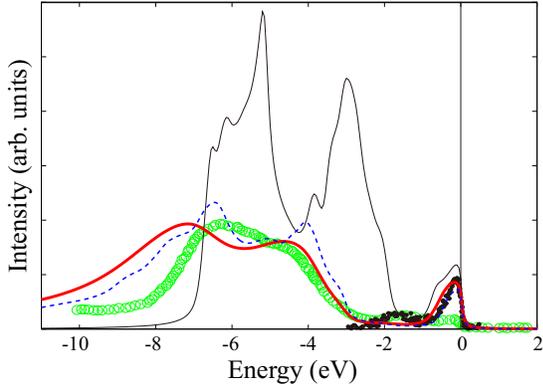}
\caption{(Color online) (a) {\it Ab initio} photo-emission spectra and experimental ones. Red-thick-solid, blue-dotted, and black-thin-solid curves are the $GW$+$C$, $GW$, and GGA results, respectively. In the experiment, light-green circles and black dots denote the photoemission spectra with the photon energies $h\nu\sim$ 60 eV (Ref.~\onlinecite{PES-SVO-60}) and $h\nu\sim$ 900 eV (Ref.~\onlinecite{PES-SVO-900}), respectively. A Lorentzian broadening of 0.05 eV is applied to the calculated spectra.}
\label{pes-SVO}
\end{center}
\end{figure}

Figure~\ref{pes-SVO} compares theoretical photo-emission spectra with experimental ones taken from Refs.\onlinecite{PES-SVO-60} and \onlinecite{PES-SVO-900}. Light-green circles and black dots are the experimental photoemission spectra with the photon energies $h\nu\sim$ 60 eV (Ref.~\onlinecite{PES-SVO-60}) and $h\nu\sim$ 900 eV (Ref.~\onlinecite{PES-SVO-900}), respectively. On the overall profile, the {\em ab initio} $GW$+$C$ spectrum reasonably reproduces the experimental results such as the relative intensity between the $t_{2g}$ and O$_p$ bands.~\cite{Gatti} On the other hand, we see that the theoretical incoherent intensity around $-1\sim-$2 eV is weaker than the experimental one.   

\begin{figure}[htbp]
\vspace{0cm}
\begin{center}
\includegraphics[width=0.4\textwidth]{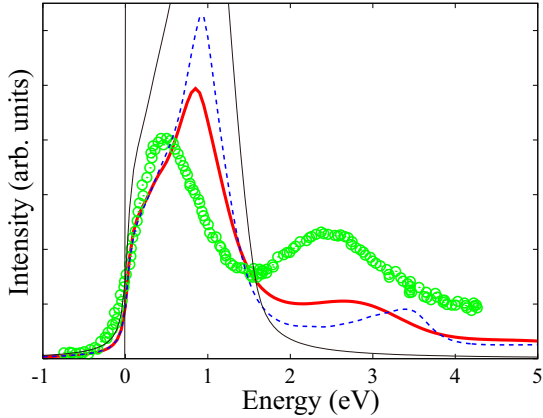}
\caption{(Color online) Comparison between {\em ab initio} spectra and experimental one (green open circles) for the unoccupied region of the $t_{2g}$ bands. The experimental data are taken from the soft x-ray absorption spectrum (Ref.~\onlinecite{XAS-SVO}). Red-thick-solid, blue-dotted, and black-thin-solid curves are the $GW$+$C$, $GW$, and GGA results, respectively. A Lorentzian broadening of 0.05 eV is applied to the calculated spectra.}
\label{xas-SVO}
\end{center}
\end{figure}

In Fig.~\ref{xas-SVO}, we compare theoretical spectra of an unoccupied region in the $t_{2g}$ bands  
\begin{eqnarray}
A(\omega>\mu)=\int_{{\rm BZ}} A({\bf k},\omega>\mu) d{\bf k} 
\end{eqnarray} 
with experimental one taken from the soft x-ray absorption spectrum (green circles) (Ref.~\onlinecite{XAS-SVO}). The effect of the low-energy plasmon excitation on the electronic structure appears to be stronger in the unoccupied region than in the occupied one.~\cite{Gatti} The {\em ab initio} $GW$+$C$ spectrum resembles the experimental spectrum more closely than the $GW$ result; the cumulant expansion reduces the quasiparticle intensity and shifts the plasmon satellite to a lower energy. We note that the theoretical spectra do not include the contribution from the $e_g$ states. Also, in SrVO$_3$, since the local-interaction effect competes with the plasmon excitation, the view of the competition of the several factors would be important for quantitative understanding, which remains to be explored. 

\section{Conclusion}\label{sec:conclusion}
We have performed {\em ab initio} $GW$ plus cumulant-expansion calculations for an organic conductor (TMTSF)$_2$PF$_6$ and a transition-metal oxide SrVO$_3$ to study the low-energy plasmon-fluctuation effect on the electronic structure. The bands around the Fermi level of these materials are isolated from the other bands, and the low-energy plasmon excitations derived from these isolated bands exist. Our calculated reflectance spectra well identify the experimental low-energy plasmon peaks. By calculating the cumulant-expanded Green's function based on the $GW$ approximation to the self-energy, we have simulated spectral functions and compare them with photoemission data. We have found agreements between them, indicating that the low-energy plasmon excitation certainly affects the low-energy electronic structure; it reduces the quasiparticle spectral weight around the Fermi level and leads to the weight transfer to the satellite parts. This effect was found to be more or less appreciable in (TMTSF)$_2$PF$_6$ than in SrVO$_3$. In particular, in (TMTSF)$_2$PF$_6$, the spectrum at the standard $GW$ level exhibits a clear plasmaron state, but considering the plasmon-fluctuation effects not treated in the standard $GW$ calculation leads to the disappearance of the state in the $GW+C$. Since the low-energy isolated-band structure is commonly found in various materials, the low-energy plasmon effect pursued in the present work can provide a basis for understanding the electronic structure of real systems. The recent progress in the photoemission experiment for the correlated materials (Refs.~\onlinecite{TMTSF-ARPES-1,TMTSF-ARPES-2,TMTSF-ARPES-3,SVO-ARPES-1,SVO-ARPES-2,Q1D-PES-1,Q1D-PES-2,Q1D-ARPES-1,Q1D-ARPES-2,Q1D-ARPES-3}) requires a concomitant progress on the theoretical side and the present {\em ab initio} many-body calculations would provide firm basis for them. 

In the present study, we have focused on the long-range correlation and treated it effectively with the cumulant-expansion method, while the short-ranged correlation, which is appropriately described by the $T$-matrix framework, is neglected. This is a future challenge which remains to be explored. 

In addition, recent photoemission spectroscopy reveals appreciable differences in the electronic structure between the bulk and thin film systems.~\cite{SVO-surface-1} The electronic structure of the surface is very sensitive to the atomic configurations at the surface~\cite{SVO-surface-2} and therefore careful analyses for the atomic structure and its effect on electronic structure are required. The {\em ab initio} calculations for the surface effect are clearly important for the deep understanding of the spectroscopy of the real materials. This is also a future challenge. 

\begin{acknowledgements}
We would like to thank Norikazu Tomita and Teppei Yoshida for useful discussions. Calculations were done at Supercomputer center at Institute for Solid State Physics, University of Tokyo. This work was supported by Grants-in-Aid for Scientific Research (No.~22740215, 22104010, 23110708, 23340095, 23510120, 25800200) from MEXT, Japan, and Consolidator Grant CORRELMAT of the European Research Council (project number 617196). 
\end{acknowledgements} 

\appendix

\section{The off-diagonal effect on the spectral function}

We briefly describe effects of band-off-diagonal terms of the self-energy on the spectral functions. With neglecting the band-off-diagonal terms, the spectral function is calculated as 
\begin{eqnarray}
A({\bf k},\omega)=\frac{1}{\pi} \sum_{\alpha} \Biggl| {\rm Im} \frac{1}{\omega-(\epsilon_{\alpha {\bf k}} + \Sigma_{\alpha {\bf k}} (\omega) + \Delta)} \Biggr|, \label{Akwdiag}
\end{eqnarray}
where the matrix element of the self-energy with respect to the KS state $|\phi_{\alpha{\bf k}}\rangle$ 
\begin{eqnarray}
\Sigma_{\alpha{\bf k}}(\omega)=\langle\phi_{\alpha{\bf k}}|\Sigma^{X}+\Sigma^{C}(\omega)-V^{xc}|\phi_{\alpha{\bf k}}\rangle   \label{Sigma} 
\end{eqnarray}
is the diagonal term of Eq.~(\ref{Sigma-od}). The $\Delta$ in Eq.~(\ref{Akwdiag}) is the energy shift to correct the mismatch of the Fermi level between the initial and final states [see Eq.~(\ref{Norm})]. 

\begin{figure}[htbp]
\vspace{0cm}
\begin{center}
\includegraphics[width=0.45\textwidth]{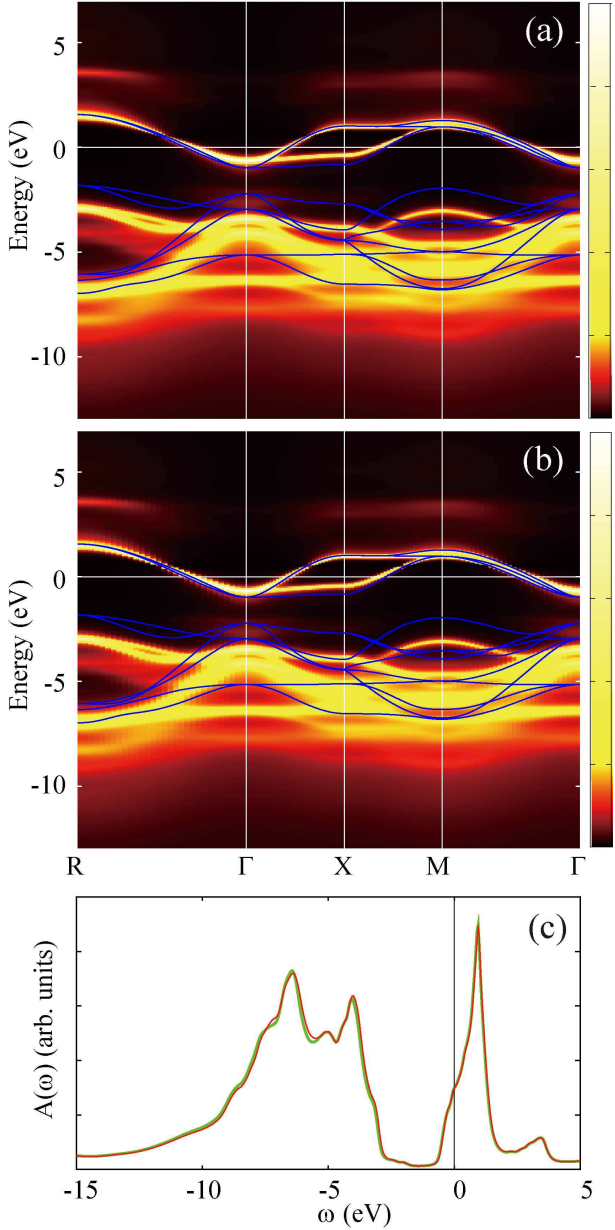}
\caption{(Color online) Spectral functions for $t_{2g}$ and O$_p$ bands of SrVO$_3$ (a) without and (b) with the band off-diagonal matrix elements of the self-energy. Blue-solid curves are the GGA result. The Fermi level is at zero energy. The colorbar is in linear scale. (c) Comparison of the density of states without (dark-red-solid curve) and with (light-green-dotted curve) the off-diagonal terms.}
\label{Akw-SVO-diag}
\end{center}
\end{figure}

Figure~\ref{Akw-SVO-diag} displays the GW spectral function of SrVO$_3$ (a) without and (b) with band off-diagonal terms of self-energy. Also, the panel (c) shows the comparison between the calculated density of states. We do not see discernible difference between the two; the band off-diagonal terms of the self-energy is negligible. This is because, in SrVO$_3$, the $d$ orbitals of the V atom are well localized and the hybridization with O$_p$ orbital is small. Similarly to SrVO$_3$, the band-off-diagonal effect on the spectral function is found to be very small in (TMTSF)$_2$PF$_6$. 


\begin{thebibliography}{999}
\bibitem{Wannier-Orbital-TMTSF} 
For (TMTSF)$_2$PF$_6$, we constructed the maximally localized Wannier orbitals for three separated energy windows, which contains HOMO, HOMO$-$1, and HOMO$-$2 bands respectively. As an initial guess for the construction of the HOMO Wannier orbital in (TMTSF)$_2$PF$_6$, we used the $p$-type Gaussians with the width 1.67 {\AA}, which is placed on the center of the central C bond of the TMTSF molecule, in a direction perpendicular to the bond. Also, an initial guess for the HOMO$-1$ Wannier orbital is a superposition of the $p$-type Gaussians with width 1.18 {\AA} put on the C atoms forming the center bond, in a direction perpendicular to the bond. In the HOMO$-2$ band, the Wannier orbital is found to be localized in only one of the two molecules in the unit cell. Therefore, we take as an initial guess for the HOMO$-2$ Wannier orbital a superposition of the $p$-type Gaussians with width 1.18 {\AA} put on the Se atoms in the molecule, in a direction perpendicular to the molecular plane.
\bibitem{Wannier-Orbital-SVO} 
For SrVO$_3$, we constructed Wannier functions from the single energy window containing V 3d and O 2p orbitals. As the initial guess, $d$- and $p$- Gaussians are set on the V and O sites, respectively.
\end{thebibliography}
\end{document}